# Economic Impact of IoT Cyber Risk - Analysing past and present to predict the future developments in IoT risk analysis and IoT cyber insurance

*Petar Radanliev, Dave De Roure* *, *Stacy Cannady, Rafael Mantilla Montalvo* [†], *Razvan Nicolescu*, *Michael Huth* ♦,

*The University of Oxford，Oxford e-Research Centre, UK, [†] Cisco Systems, USA, ♦ Imperial College London, UK

## Abstract

This paper is focused on mapping the current evolution of Internet of Things (IoT) and its associated cyber risks for the Industry 4.0 (I4.0) sector. We report the results of a qualitative empirical study that correlates academic literature with 14 - I4.0 frameworks and initiatives. We apply the grounded theory approach to synthesise the findings from our literature review, to compare the cyber security frameworks and cyber security quantitative impact assessment models, with the world leading I4.0 technological trends. From the findings, we build a new impact assessment model of IoT cyber risk in Industry 4.0. We therefore advance the efforts of integrating standards and governance into Industry 4.0 and offer a better understanding of economics impact assessment models for I4.0.

## 1  Introduction

The evolution of IoT represents multiple categories of cyber-physical systems, integrating technologies related to smart grids, smart homes, intelligent transportation, manufacturing and supply chain and smart cities, to name a few. Such new technologies come with new types of risks that existing risk assessment/management methods are not designed to anticipate or predict. Safeguarding an IoT deployment IoT, while simultaneously harnessing its economic value, requires systematic consideration of multiple factors, including: privacy, ethics, trust, reliability, acceptability and security. Such a systematic approach would go far to ensure the integrity, confidentiality, and availability of the data contained in IoT devices and services. Cyber security has been recognised as a critical national policy issue. by many countries Economic impact of cyber risk and cyber security importance is growing as the integration of IoT connected devices into smart manufacturing and supply, cities, intelligent transport systems, smart grids and more aspects of modern life, including banking, finance, autonomous cars and personal medical devices. Cyber-attacks are increasing in frequency, and the and increasingly target IoT devices (for example the Mirai botnet). The severity of future attacks could be much greater than what has been observed to date.

A critical question for government policy and for private sector business strategies for IoT connected products, platforms and services is the sufficiency of cyber security to minimize cyber risk that accompanies IoT deployments. This answer must be partially addressed by economic analysis, such as cost and frequency analysis of cyber-attacks. Such analysis would complement the process of building frameworks and methodologies for mitigating the economic impact of cyber risk of commercial use of deployments of IoT connected products and services.

The research problem investigated in this paper is the present lack of standardised methodology that would measure the cost and probabilities of cyber-attacks in specific IoT related verticals (ex. connected spaces or commercial and industrial IoT equipment) and the economic impact (IoT product, service or platform related) of such cyber risk. As a result, the growth of the IoT cyber risk finance and insurance markets are lacking empirical data to construct actuarial tables. Despite the development of models related to the impact of cyber risk, there is a lack of such models related to specific IoT verticals. Hence, banks and insurers are unable to price IoT cyber risk with the same precision as in traditional insurance lines. Even more concerning, the current macroeconomic costs estimates of cyber-attacks related to IoT products, services and platforms are entirely speculative. The approach by 'early adopters' that IoT products are 'secure by default' could be somewhat misleading. Even governments advocate security standards ex. standards like ISA 99, or C2M2 [1], [2] that accept that the truth on the ground is that IoT devices are unable to secure themselves, so the logical placement of security capability is in the communications network.

The research methodology in this paper proposes combining the Cyber VaR, NIST and FAIR frameworks to build a new model for calculating the economic impact of IoT cyber risk. There is a limited research on the economic impact of cyber risk. There is even less research on the economic impact related to cyber risks from different IoT verticals. The economic impact of IoT related cyber risks in present time are assessed by applying methodologies established before the development of IoT verticals (ex. automated, digital, social machines, cyber-physical and coupled systems). Present day critical infrastructure systems are far more complex, creating new risks for failures. Further, risk in an IoT deployment might extend to many entities. A interruption in services delivered by a smart grid or smart city would impact many businesses, agencies and individuals. For example, failure in MY IoT deployment might cost millions due to interrupted services. This creates the rationale that a new impact model and assessment methodology are needed that would anticipate economic impact of cyber risks and benefits from the IoT ecosystem. This research would build upon existing cyber risk models (e.g. VaR, Cyber VaR). The research aim is to develop a robust economic model to estimate the economic impact in IoT verticals (ex. communications network, or critical infrastructure).

### Genesis of IoT

The IoT term was created in 1999 [3] and the first IoT principles were published shortly after in the book 'When Things Start to Think' [4]. According to Gartner's IT Hype Cycle, the IoT market adoption will take 5-10 years, as of 2012 [5].

### Research rationale

Cyber risk in the IoT is increasing at an alarming rate and cyber security is of increasing relevance to early adopters for harnessing economic value from the IoT, without exposing critical infrastructure to cyber risks. Some of the technologies (not all) that are used every day are (at present) not connected with the internet, such as: gas meters, house lights, healthcare devices, water distribution systems, cars and other road transport vehicles. However, such devices are increasingly becoming digitally connected and communicating through mobile (or wireless) networks, e.g. M2M. Some examples include connected spaces, smart meters and autonomous cars. Ultimately IoT may revolutionise our business ecosystem. This evolution is triggered by a number of factors and forces. Some include: objects connected to the IoT can reduce costs through use of the data they collect, create business opportunities, and can promote new services. IoT products and services are disadvantaged compared to non-connected devices, because of the concern over cyber risks. This is similar to the story that played out during the emergence of cloud computing. It seems likely that customer concern will drive new opportunities for promoting cyber security that could lead to reducing this gap in competitiveness triggered by higher cost and fears of cyber-attacks.

The growth of the IoT market (ex. in the critical infrastructure vertical) could increase significantly if policymakers have the methodology to assess, predict, analyse and address the economic risks of IoT related cyber-attacks in the communications network. Without the appropriate risk assessment methodology, the likelihood of serious economic impact due to attack, can only be determined by subjective assessment. Connecting the economic impact of different IoT verticals cyber risk to critical infrastructure through impact models, can provide feedback sensors and real time data mechanisms. This would assist and enable industry and policymakers to visualise the problem and address the economic risk created by IoT related cyber-attacks.

### New Theoretical Frontiers

New theoretical model that integrates cyber risks from the physical and cyber subsystems is necessary. The new theoretical model needs to provide an overall understanding of the design, development, and evolution of IoT cyber risks. The model needs to integrate theories of IoT, control of physical systems, and the interaction between the physical and the digital worlds.

## 2 Literature review

### Economic value of IoT digital infrastructure

According to a 2013 Cyber Power Index [6], The United Kingdom has been ranked as the overall global superpower followed by the United States. However, according to the same report, the analysis of industry application of digital infrastructure in key sectors (Smart Grids, E-Health, E-Commerce, Intelligent Transportation and E-Government), The United Kingdom drops much lower to the 5th place and United States on the 3rd place of the index. It seems that the UK and US are strongly protected to withstand digital infrastructure cyber-attacks, which is crucial in developing digital economy [7]. But the UK and US seem to be lagging behind in terms of capabilities to capitalise on the new digital era. This lagging behind in the harnessing of economic value from digital infrastructure could be caused by the barriers to adoption of smart manufacturing technologies (such as cost), especially for small enterprises [8]. New infrastructure for smart manufacturing technology would create large savings for manufacturers, in the US the savings are estimated to $57.4 billion annually [8]. This could improve the harnessing of economic value from digital infrastructure, but the concerns about the economic impact of IoT cyber risk would remain, especially in cyber risk insurance policies for SME's. For example, ICT cyber insurance either gives genuine protection, or it offers more of a consulting relationship where the insurance provider offers initial training and measures, and will come in after an attack to assist in recovery. These approaches do not seem to be in the interest of SMEs because the cyber

insurance cab be void (e.g. if the insurance broker finds out that a specific software update was not done by midnight of a certain date) and the recovery usually comes at a premium price. As such, current cyber insurance policies do not seem to considerably help with making ICT systems more resilient against cyberattacks. If cyber insurance companies could predict with precision the maximum economic impact, this could enable the insurance companies to provide more comprehensive policies which would help SME's protect against cyber risk impact that exceeds their individual risk impact tolerance level.

## Economic impact of IoT cyber crime

Cyber risk has not been clearly quantified through historical measures because of the risk environment is changing fast [9]. The common figure stated is a loss of $1 trillion to cybercrime, but estimates range from: 300bn and $1tn [10], $400bn to over $575bn [9], or $400bn to over $2tn [11]. The difference in these figures shows that the numbers are rough estimates at best, and the real economic impact of cyber risk remains unknown [11]. The main difficulties in calculating the economic impact of cyber risk are the lack of suitable data and the lack of universal standardised framework to assess cyber risk [12]. Adding to these, there is the need to quantify accumulated risk on a shared technology platform (such as cloud computing) and hyper-connectivity in the digital supply chain [13]. Analysing the economic impact of cyber risk is also complicated because of the impact on brand reputation, the cost of downtime, legal liability, cost of intellectual property loss, and many other variables. Merely the media coverage of cyber risk has created such significant economic impact that managing risk has become 'imperative' [10].

## Economic impact IoT data ownership

In terms of data ownership, data privacy and Economic lifespan of digital assets, it has already been established that digital assets can outlive humans [14], triggering the question of data ownership after end of data owners' life. Adding to this argument, a large quantity of low-quality or duplicated data are never deleted, creating 'data pollution'. Such complex topics triggers the question of do we need to set a 'self-deletion' phase. Some studies have simplified the topic with the assumption of a limited economic lifespans for all classes of digital assets [13]. Because human society is an event driven system, where digital abstractions of the physical world have a lifespan.

## Economic impact of IoT

IoT is essential for future economic competitiveness, but technological innovations are necessary for harnessing the economic value [15]. Maximising the economic impact of IoT should contain: extreme-yield agriculture [16] supported by energy-aware buildings and cities [7], physical critical infrastructure with preventive maintenance, and self-correcting cyber-physical systems [15], [16]. On the other hand, the economic impact of IoT cyber risk can be quite damaging. The electric power grid represents one of the largest complex interconnected networks, and under stressed conditions, even a single failure can trigger complex cascading effects, creating wide-spread failure and blackouts, [16]. Distributed energy resource technologies such as wind power, create additional stress and vulnerabilities [7], [15], [16].

## Economic Impact of Cyber Risk from the Internet of Things

The world is experiencing the fourth industrial revolution [16]–[18], where the IoT real-time enabled platforms [7] represents the foundation for digital industry [17], [19]. Digital industry would be supported with more intelligent, resilient and interconnected manufacturing equipment [7], [20], [15]. The integration of artificial intelligence (AI), machine learning, the cloud, and IoT will create systems of machines capable of interacting with humans [7], [19]. The application of behavioural economics into these systems of machines [21] already enables market speculation on human behaviour [22] and even neuromarketing [23] to determine consumer purchasing behaviour. We can expect to see autonomous machines adopting the use of this methods to predetermine human behaviour [19].

Technologies that would enable the integration of IoT in the digital industry include software defined networks [24] and software defined storage [25]. The foundations that IoT and CPS industrial integration are built upon are protocols and enterprise grade cloud hosting (Carruthers, 2016); AI, machine learning, and data analytics [26]; and mesh networks and peer-to-peer connectivity [27]. IoT transforms the sensory and control cyber physical systems, creating security and risk management vulnerabilities due to many factors, including complexity of the deployment, uncertainty of the inventory in the deployment, the access points of the deployment to the Internet and from integrating less secured or unsecured systems, triggering into the deployment. This s many questions on risk management and liability for breaches or damages [19].

Cyber risk mitigation modelling requires:

- A management strategy for: espionage, theft, or terrorist attacks, which in effect requires electronic and physical security [16], [15].

- Insider threats must also be covered, including interception and analysis of non-communications electromagnetic radiations [9].

- A cyber risk mitigation model also requires information assurance, data security and protection for data in transit,

from physical and electronic domains and storage facilities [7], [9], [28].

- A cyber risk mitigation model requires anti-counterfeit and supply chain risk management to counteract components introduced in the supply chain, modified from its original design to enable a disruption or an unauthorised function [9], [29].

- Limiting the source code access to crucial personal provides software assurance and application security is necessary for eliminating deliberate flaws and vulnerabilities [16].

- A cyber risk mitigation model should be supported with forensics, prognostics, and recovery plans, for analysis of cyber-attacks and for coordination with agencies responsible to identify external cyber-attack vectors [9]. Internal track and trace network process can assist in determining and prevent the existence of weaknesses in the logistics security controls [9].

- Anti-malicious and anti-tamper system process is needed to prevent vulnerabilities identified through reverse engineering attacks [9], emphasising the need for security and privacy [16]. To prevent continuation of cyber-attacks, information sharing and reporting, fast cyber-attack reporting and shared database resources should also be developed ([9], [17]).

## 3 Research methodology

This section outlines the research methodology applied in the research. The section starts with detailing the models applied and adapted. Then the complexities of designing a new impact assessment model are discussed. Finally, the early models are compared with most research modelling approaches to define the rationale for the research methodology applied.

### Economic impact frameworks and models

The Cyber Value-at-Risk (CyVaR) framework has been promoted for standardisation of language, models and methods [30] which has been further developed by Deloitte (2016). This framework represents the first attempt to understand the economic impact of cyber risk for individual organisations [12] The first unifying economic framework encompassing the cross-disciplinary field of 'Cybernomics' proposed measurement units for cyber risk [13]. Multidisciplinary methodologies are applied, along with established risk measurement methods to define individual risk units: e.g. MicroMort (MM) for measuring medical risk, Value-at-Risk (VaR) for measuring market risk for measuring cyber risk [13]. The main weakness of this framework is that it has not been tested or validated with real data. It has taken years to validate VaR and decades to validate MM due to the time required for data collection. Other cyber value analysis methods have advanced to calculate the cost of different cyber-attack types [31], but the same problem with lack of data to validate the model persists. This lack of data has motivated the development of a proof of concept method [12] that is based on data assumptions. The weakness in this approach is that economic impact is calculated on organisations' 'stand-alone' cyber risk, because data assumptions can only be made on individual cases. However, Business impact for the same risk can vary widely between companies based on the specific circumstances of each company. Furthermore, that approach ignores the correlation effect of organisations sharing infrastructure and information, and by default, sharing cyber risk exposure. Cyber risk exists in multiple physical, information, cognitive, and social domains, (software, hardware, firmware, adjacent systems, energy supplies, supply chains) and the economic impact is related to these closely interconnected systems. This close interconnection of disparate systems increases the probability of 'cascading impacts' [9]. This is of great concern especially in sharing cyber risk in critical infrastructure [12], because critical infrastructure is vital for a strong digital economy [16].

### Complexities in building economic impact theoretical model

There are multiple problems in building one theoretical model that would rule all of the complexities discussed. There are additional complexities that are almost impossible to quantify. For example, in information assets such as intellectual property of digital information, the future value is lost regardless of early detection [12]. Therefore, the economic value of digital assets has to reflect their economic functions first before their value can be properly assigned [13].

Table 1 lists a number of cyber risk management methodologies as used or proposed in industry and academia.

| Qualitative Methods |
|---|
| 1) The IT Infrastructure Library (ITIL) |
| 2) Control Objectives for Information and Related Technology (COBIT) |
| 3) ISO/IEC 27005:2011 |
| 4) Information Security Forum (ISF) Simplified Process for Risk Identification (SPRINT) and Simple to Apply Risk Analysis (SARA) |
| 5) Operational Critical Threat and Vulnerability Evaluation (OCTAVE) |
| 6) NIST Special Publication 800-53 |
| 7) NIST Special Publication 800-37 |
| 8) ISO/IEC 31000:2009 |
| 9) Consultative, Objective and Bi-functional Risk Analysis (COBRA) |
| 10) Construct a platform for Risk Analysis of Security Critical Systems (CORAS) |
| 11) Business Process: Information Risk Management (BPIRM) |
| Quantitative Methods |

12) Information Security Risk Analysis Method (ISRAM)
13) Central computer and Telecommunication Agency Risk Analysis and Management Method (CRAMM)
14) BSI Guide- RuSecure- Based on
15) BS7799 Standard
16) Cost-Of-Risk Analysis (CORA)

Existing cyber risk frameworks and methodologies are constrained by a number of limitations. Cyber risk assessment frameworks are based on security control domains and assess security posture, but are not effective in assessing high risk loss scenarios developed around critical digital assets [13]. Furthermore, cyber risk assessment methodologies have created an inconsistency in measuring cyber risk, because of the absence of a common point of reference [13].

## Comparison of early and more recent models on the economic impact of cyber risk

Earlier literature suggested methods based on Return on Investment (ROI) and Net Present Value (NPV), have been proposed to assess the information security investment, that include broad set of criteria, including 'economics of privacy' [32], 'optimal amount to invest' [33], 'risk averseness' [34], but these methods are not validated with real data. In addition, cyber risk covers more elements than information security financial cost, and a method is needed that would integrate cyber risk directly with economics [13]. Because the motivation for cyber risk can be different than purely financial (ex. espionage), and yet still creating economic impact. Therefore, the impact should be calculated in terms of average and in the most severe scenario [12].

To make such calculations with a reasonable precision of the impact assessment, different modelling approaches need to be integrated in a new and more reliable economic impact assessment model. This research proposes a design of such model for calculating the economic impact of IoT cyber risks, by integrating the CyVaR with the MM model and the recommendations from earlier models.

## 4 The model

We need a reliable model for costing cybercrime [35] and the first step in developing a costing model for IoT cyber risk, is to determine the cybercrime units of costings. To determine the risk of cybercrime, we refer to established methods for calculating risk.

Risk = Likelihood × Consequences, and cyber-risk can be defined as a function of:

$R = \{s_i, p_i, x_i\}, i = 1, 2, \ldots, N,$

$R$ – risk; $s$ – the description of a scenario (undesirable event); $p$ – the probability of a scenario; $x$ – the measure of consequences or damage caused by a scenario; $N$ – the number of possible scenarios that may cause damage to a system.

To build a model for calculating the impact of IoT cyber risk, we need to combine established risk models [13], such as MicroMort (MM) and Value-at-Risk (VaR) for measuring market risk and adapt a new cyber risk units for IoT MicroMort (IoTMM) and IoT MicroMort2 (IoTMM2) as the value of reducing the risk by a given IoTMM.

The economic functions of IoT assets requires an International IoT Asset Classification (IIoTAC). The term is chosen to be compliant with the proposed International Digital Asset Classification (IDAC) [13].

IoT digital assets can be categorised as: (1) IoT core value assets (IoTCA), where digital assets which are directly part of goods or services that T profits from; (1a) IoT digitised assets (IoTDA), where goods and services digitised from traditional goods and services; (1b) IoT assets born digital, representing things and services that are intrinsically digital; and (2) IoT operational assets (IoTOA), representing assets that support the creation, consumption and distribution of IoT goods and service.

Thing's (T) IoT composition can be described by the ratio of its core value assets to operational assets: $CA:OA = \{c_i, p_i\} : \{o_j, q_j\}$ $i=1,2,\ldots,N_c$, $j=1,2,\ldots,N_o$ where

IoTCA – T's core value assets; IoTOA – T's operational assets; $c$ – a type of asset listed in IDAC which is of core value to T; $p$ – T's core digital asset $c$; $o$ – a type of asset listed in IDAC which is of operational value to T; $q$ – T's operational asset $o$; $N_c$ – the number of core value assets in T; $N_o$ – the number of operational assets in T.

By using the same formula, T's DA (digitised assets) to AD (assets born digital) ratio can also be calculated. T's digital value composition describes its nature of innovation, e.g. traditional goods have a high OA:CA ratio, while software has a high CA:OA ratio and a high AD:DA ratio. Other valuation parameters are: Intrinsic value of IoT digital asset can be determined through fundamental analysis without reference to its market value. Market value of IoT digital asset is the price at which the digital valuable would trade in a competitive market. Subjective value of IoT digital asset is determined by the importance the T places on it.

Following these valuation parameters, the value of (1a) IoT assets is directly converted from their physical equivalents. The value of (1b) IoT assets requires their own valuation analyses. (2) IoT assets can be valued with Business Impact Analysis (BIA). According to this formula of the existing economic theory of value to digital asset, the T's total digital value can be calculated as:

$$V = \sum_{i=1}^{Nc} cv_i + \sum_{j=1}^{No} ov_j$$

where:

V – total digital value of T; cv – value of core value asset c of T; ov – value of operational asset o of T; Nc – the number of core value assets in T; No – the number of operational assets in T.

This valuation requires Key IoT Cyber Risk Factors (KIoTCRF) correlated with a T's risk profile. Established Key Cyber Risk Factors (KCRF) risk categorisations [13] can be adopted to IoT, where: Technological factors are related to the usage of technology. Non-technological factors are related to: people, process, socio- economic, geo-political factors. Inherent factors are related to T's nature of business, industry, core operations, goods and services. Control factors represent T's control effectiveness against cyber loss. Therefore, the T's residual cyber risk can be calculated as:   Residual cyber risk = inherent risk ÷ control effectiveness. This valuation allows for MM to be applied to define cyber risk units for class D assets and to define IoT MicroMortD (IoTMMD) for a given class D digital assets as 1 in a million probability of its digital death, where the value of 1 IoTMMD is the amount of money T is willing to pay to reduce 1 IoTMMD for its class D assets.

Since IoT residual risk IoTMM is not statistically available, when it becomes statistically available for various types of IoT assets, it could be aggregated with asset values to generate a cyber VaR curve, representing T's residual cyber risk:

$$VaR = \sum_{i=1}^{n} V_i f_{Di},$$

To compute the cyber VaR curve, historical simulation and Monte Carlo simulation can be used, where VaR is Value-at-Risk for all IoT digital assets of T; T's digital asset inventory D = {D1, D2, ..., Dn}; the value of each asset V = {V1, V2, ..., Vn}; and fDi is the amount of residual risk Di is exposed to, measured in IoTMMD is. Monte Carlo can generate a large number of paths using repeated random sampling to produce a probability distribution. In this scenario, the risk measure IoTMM2 can be defined as a 12-month IoTMM2 VaR representing the loss limit T can afford from cyber incidents. Where IoTMM2 is the cost T is willing to pay to reduce its IoTMM2 by 1% for the same loss limit. The VaR can be calculated for 12 months to represents cyber risk exposure over one financial year, required for budget planning in ERM frameworks.

The proposed valuation depends on advanced data analytics, capable to support a trajectory of exponential growth. We have the advantage of storing and processing large datasets, hence the main obstacle is not the lack of capabilities to compute datasets, but to break down non-technological barriers and establish a wide range of data points in the proposed categories.

It may take years or decades to validate the economic impact of IoT cyber risk, because of the time required for data collection. However, it is important to set the categories in order for the data collection to be performed in a structured manner.

## 5 Applying the proposed model for IoT MicroMort calculations

To test, validate and verify the findings of the new model, (a) the IoTMM for 2017 is calculated; and (b) for 2020 is forecasted, from the following data. There are estimated 378 Million Devices Potentially Vulnerable to Hacking in 2017 out of 8.4 billion connected things [36]. These numbers emerged from the BullGuard's IoT Scanner, where 310,000 users scanned their network for vulnerabilities and 4.5 percent (nearly 14,000 devices), were reported as 'could be easily hacked'. This data is combined with Garner report that 8.4 billion connected things will be in use worldwide in 2017 [37]. To forecast the IoTMM for 2020, the forecasted data is used from the same report showing that the number of IoT connected devices will reach 20.4 billion by 2020, with more than 900 million potentially vulnerable devices by 2020.

Therefore, (a) the IoTMM for 2017 is calculated as 0.045

and (b) the IoTMM for 2020 is calculated as 0.044

The next step is to calculate the enterprises 'willingness to pay' to reduce 1 IoTMM. This is representative of the cost sum for an enterprise to accept a one-in-a-million IoTMM, or the cost sum that enterprise might be willing to pay to avoid a one-in-a-million chance of IoTMM. For the purposes of testing this model, we could apply a nominal Value of a Statistical Life (VSL) or the Value for Preventing a Fatality (VPF) to evaluate the cost-effectiveness of expenditure on cyber security. The IoT security spending is estimated to increase to $840.5 million in 2020 [38]. This would IoT market value of 1 IoTMM in 2020 as $840.5. However, it is important to understand what does the value of 1 IoTMM represent in this scenario. We can explain this with an example, e.g. each T in a sample of 100,000 T's willingness to pay for a reduction in their individual IoT risk of 1 in 100,000, or 0.001%, over the next year. Since this reduction in risk would mean that we would expect one fewer IoTMM among the sample of 100,000 T's over the next year on average. Supposing that the answer was $840.5, then the total dollar amount that the group would be willing to pay to save one statistical life in a year would be $840.5 per T × 100,000 T's, or $84,050,000 million. This is a very generic estimate that cannot be used by governments as guidance point for creating standards and governance. Calculating the IoTMM for 8.4 billion connected things would result with a number far greater than the estimated IoT security spending of $840.5 million in 2020. Unfortunately, we have no data as to how the experts estimated the IoT security spending, and the utility functions in such estimates are often not linear. Therefore, the economic value of 1 IoTMM does not represent a precise calculation of the value and risk. It represents more of a guidance point to show that as more IoT devices become connected, their cyber security is not competitively priced, which increases the risk, and we need to be aware that we have no precise calculation of the IoT cyber risk, or cyber risk in general.

Enterprises can obtain a valuation more precise to their T's by assessing the previously described valuation formula where T's digital asset inventory D = {D1, D2, ..., Dn}; combined with the value of each asset V = {V1, V2, ..., Vn}; and fDi is the amount of residual risk Di is exposed to, measured in IoTMMD is. Resulting with the calculation of the value of 1 IoTMMD in 2020 as the amount of money T is willing to pay to reduce 1 IoTMMD for its class D assets, valued with:

$$V = \sum_{i=1}^{N_c} cv_i + \sum_{j=1}^{N_o} ov_j$$

## 6 Discussion

The figures we are applying are just to verify the new model. Since there is no International IoT Asset Classification (IIoTAC) and no established Key IoT Cyber Risk Factors (KIoTCRF), the calculations of the new model serve just to verify the new model. After the establishment of IIoTAC and KIoTCRF, the new model could be applied to calculate more precise 'willingness to pay' that T is willing to pay to reduce 1 IoTMMD.

We need to mention that the local linearity of the utility curve means that the MicroMort is useful for small incremental risks and rewards, not necessarily for large risks. Therefore, the IoTMM is not an ideal measure to calculate the IoT risk. Instead, IoTMM is better placed to measure for a given T willingness to pay to reduce 1 IoTMMD for its class D assets.

Finally, we need to discuss the lack of IoT data. For example, the latest forecast from Gartner Inc. says worldwide information security spending will reach $86.4 billion (USD) in 2017 and $93 billion in 2018. That forecast doesn't cover the IoT, ICS (Industrial Control Systems) and IIoT (Industrial Internet of Things) security [39]. Given the lack of data on IoT cyber risk, cyber loss, or profits from different IoT vectors, it is extremely difficult to conduct IoT cyber risk analysis and argue on the soundness of the analysis. Since the cyber insurance is in its infancy, insurance companies have not mastered the valuation of cyber risk in general. For example, Target was insured for $100 millions of cyber risk in 2017, and suffered over $450 millions of loss, with estimated to total at $1 billion by the end of 2017 [40]. This example clearly states that cyber insurance needs a lot more data to calculate, correlated and transfer risk with an acceptable degree of certainty. While general cyber risk cannot be calculated, the emergence of IoT has created new IoT risk vectors that are not at all defined in the cyber insurance policies.

## 7 Conclusion

The findings from this research lead to the conclusion that there many challenges in understanding the types and nature of cyber risk and their dependencies/interactions in this new space. This paper informs on how one may assess economic impact with mathematical formalisms.

The multiple complexities explained in the study, in terms of calculating the economic impact of IoT cyber risk, also lead to the conclusion that impact can only be assessed with new risk metrics, and a new valuation method specific for the new risk metrics, combined with new regulatory framework and standardisation IoT data bases with new risk vectors as defined in the form of International IoT Asset Classification (IIoTAC) and Key IoT Cyber Risk Factors (KIoTCRF).

This paper presents new risk metrics, by adapting established methods for calculating risks and uncertainties, and identifies some specific grand challenges for calculating the economic impact of IoT cyber risk. The paper combined common basic terminology, common approaches and incorporated existing standards into a new model for calculating the economic impact of IoT cyber risk**.**

This work was supported by the UK EPSRC with project [grant number EP/N02334X/1 and EP/N023013/1] and by the Cisco Research Centre [grant number 2017-169701 (3696)].